\documentclass[a4paper,11pt]{article}
\usepackage{pos}
\usepackage{verbatim,ulem,amsmath}

\newcommand{\mcO}{{\mathcal O}}
\newcommand{\wmcO}{\widetilde{\mcO}}
\newcommand{\GF}{\mathrm{GF}}
\newcommand{\MSbar}{$\overline{\mathrm{MS}}${}}
\newcommand{\ms}{\overline{\mathrm{MS}}}
\newcommand{\UV}{\mathrm{UV}}
\newcommand{\IR}{\mathrm{IR}}
\newcommand{\bx}{{\bf x}}
\newcommand{\wV}{\widetilde{V}}

\title{A novel nonperturbative renormalization scheme for local operators}
\ShortTitle{Novel renormalization scheme}

\author[a]{Anna Hasenfratz}
\author*[b,c]{Christopher J.~Monahan}
\author[d]{Matthew D.~Rizik}
\author[d]{Andrea Shindler}
\author[e]{Oliver Witzel}

\affiliation[a]{Department of Physics, University of Colorado, Boulder, CO 80309, USA}

\affiliation[b]{William \& Mary, Williamsburg, VA 23187, USA}
\affiliation[c]{Thomas Jefferson National Accelerator Facility, Newport News, VA 23606, USA}

\affiliation[d]{FRIB \& Physics Department,
Michigan State University, East Lansing, MI 48824, USA}

\affiliation[e]{Center for Particle Physics Siegen, Theoretische Physik 1,\\
  Naturwissenschaftlich-Technische Fakult\"at, Universit\"at Siegen, 57068 Siegen, Germany}

\emailAdd{anna.hasenfratz@colorado.edu}
\emailAdd{cjmonahan@wm.edu}
\emailAdd{rizik@nscl.msu.edu}
\emailAdd{shindler@frib.msu.edu}
\emailAdd{oliver.witzel@uni-siegen.de}

\abstract{The gradient flow exponentially suppresses ultraviolet field fluctuations and removes ultraviolet divergences (up to a multiplicative fermionic wavefunction renormalization). It can be used to describe real-space Wilsonian renormalization group transformations and determine the corresponding beta function. We propose a new nonperturbative renormalization scheme for local composite fermionic operators that uses the gradient flow and is amenable to lattice QCD calculations. We present preliminary nonperturbative results for the running of quark bilinear operators in this scheme and outline the calculation of perturbative matching to the  \MSbar{} scheme.}

\FullConference{%
 The 38th International Symposium on Lattice Field Theory, LATTICE2021
  26th-30th July, 2021
  Zoom/Gather@Massachusetts Institute of Technology
}

\begin{document}
\maketitle

\section{Introduction}
Quantum chromodynamics (QCD), the gauge theory of the strong nuclear force, is strongly coupled at hadronic energy scales. Lattice QCD provides a numerical approach to studying the nonperturbative features of QCD by discretizing spacetime and evaluating the path integral stochastically with Markov chain Monte Carlo methods. The discretized spacetime provides a gauge-invariant ultraviolet regulator of QCD, and taking the continuum limit corresponds to removing this ultraviolet regulator. This, in turn, requires that both the lattice QCD action and all probe operators are renormalized. For QCD at hadronic energy scales, this requires a nonperturbative renormalization procedure. Ideally, a fully nonperturbative renormalization scheme for lattice QCD should preserve gauge-invariance and be amenable to numerical implementation on the lattice and calculable in perturbation theory at high energies. Moreover, a method for extracting the nonperturbative energy dependence that allows matching results in both low- and high-energy regimes is necessary. We propose a new renormalization procedure that satisfies these requirements and study its application to quark bilinear operators.

This new approach allows for the nonperturbative determination of the anomalous dimension of quark bilinear operators through the real-space renormalization group (RG), implemented using the gradient flow \cite{Narayanan:2006rf,Luscher:2009eq,Luscher:2010iy}. The gradient flow is not an RG transformation as it lacks the crucial step of rescaling (coarse graining). However, as Ref.~\cite{Carosso:2018bmz} outlines, a coarse graining step can be implemented at the level of expectation values, transforming the gradient flow to a complete real-space RG transformation.   There are three steps. First, we calculate the nonperturbative renormalization parameter $Z_\mcO^{\GF}(\mu,a)$ at a low-energy scale, through a ratio of matrix elements of gradient-flowed operator $\mcO$.
Then, we evaluate 
the nonperturbative anomalous dimension, $\gamma_\mcO$ by computing the numerical logarithmic derivative of $Z_\mcO^{\GF}(\mu,a)$. Finally we match the nonperturbative gradient flow scheme to the \MSbar-scheme (or similar) at high energy, using a short flow time expansion (SFTE) applied to our gradient flow scheme.

There are two widely-used nonperturbative renormalization schemes in lattice QCD, the Schr\"odinger functional \cite{Luscher:1992an,Sint:1993un}, and regularization-independent momentum subtraction (RI/MOM) schemes \cite{Martinelli:1994ty}. On the one hand, the Schr\"odinger functional approach is gauge-invariant and provides a natural definition of a finite volume scheme that enables the extraction of the nonperturbative  running of operators, but at the cost of introducing additional analytic  complications. On the other hand, RI/MOM schemes break gauge invariance, which introduces computational challenges for the gauge sector, and do not provide a natural step-scaling procedure to match low and high energy regimes. A previous proposal to use the gradient flow to define a renormalization scheme for local fermionic operators was limited to a finite volume step-scaling procedure for quenched QCD \cite{Monahan:2013lwa}.

Our proposed  scheme  is gauge invariant and defines anomalous dimensions nonperturbatively.  However, the smearing radius of the flow must be small compared to other length scales in the calculation, which may lead to a ``window problem''. Moreover, the flow introduces additional diagrams in perturbative matching calculations. We note that these diagrams can be computed in continuum perturbation theory, avoiding the need for lattice perturbation theory. 

Here we provide a proof-of-principle calculation of the renormalization factors and anomalous dimensions of quark bilinear operators. We anticipate that our renormalization scheme can be implemented for a wide range of local operators in lattice QCD, from flavor-changing weak currents and four-quark operators, to twist-2 operators relevant to hadron structure calculations. Calculations for further quantities are discussed in \cite{Rizik:2020naq,Harlander:2020duo,Kim:2021qae,Mereghetti:2021nkt,Harlander:2021esn}. Future work will be required to identify the region of applicability, quantify the associated systematic uncertainties, and assess the viability of our scheme for different local operators.

\section{Renormalization procedure}

Gradient flow transformations can describe real-space Wilsonian RG transformations \cite{Carosso:2018bmz,Makino:2018rys,Abe:2018zdc,Sonoda:2019ibh,Miyakawa:2021hcx,Sonoda:2020vut}, allowing the determination of the RG $\beta$ function of the running coupling constant \cite{Hasenfratz:2019hpg,Hasenfratz:2019puu,Aoki:2014dxa,Fodor:2017die,Peterson:2021lvb} and the anomalous dimension of composite fermion operators \cite{Carosso:2018bmz}. 

Real space RG transformations are constructed from blocked fields, defined over a ``coarse-grained" subset of the lattice and constructed from the original fields, which are then integrated out. Blocking transformations leave the partition function of the theory invariant, but both the parameters of the action and expectation values of operators  change. Typically, these changes have a complicated analytical form, but in the vicinity of a fixed point or renormalized trajectory the analytic structure is well understood.

The gradient flow transformation in real space averages the field variables over a region proportional to $\sqrt{2Dt}$  in a $D$-dimensional spacetime~\cite{Luscher:2009eq,Luscher:2010iy,Narayanan:2006rf}. The flowed fields are therefore a natural choice for blocked field variables. Moreover, flowed operators provide probes of unflowed composite operators. So it is natural to consider correlation functions between blocked and unblocked, flowed and unflowed operators. The gradient flow transformation does not change the lattice spacing, it has to be supplemented by a coarse graining step to define an RG transformation. Coarse graining can be applied to expectation values. That way the gradient flow describes an  RG transformation
 if the gradient flow flow-time $t$ and RG scale change $b$ are related by $b \propto \sqrt{t/a^2}$, with $\sqrt{t/a^2} \gg 1$ \cite{Carosso:2018bmz}. 

Calculating expectation values in RG flowed systems does not require simulations with the flowed action. It is sufficient to calculate expectation values of flowed operators on ensembles generated with the original action. This observation greatly simplifies the numerical task of real space RG \cite{Swendsen:1979gn}.
Nevertheless, it is worth emphasizing that the gradient flow is not an RG transformation. The RG blocked action has fixed points, while the flowed action flows to a saddle in the action parameter space. Moreover,  identifying flowed fields with blocked fields does not allow the iteration of the RG transformation, because the blocked fields from a second blocking step cannot be expressed as simple flowed variables.

In the large flow-time limit the RG flow tracks the renormalized trajectory and expectation values of scaling operators describe continuum physics. 
In particular, the operator $t^2\langle E \rangle_t$, where $\langle E \rangle_t$ is the  energy density of the gauge fields at flow time $t$, is dimensionless and can be used to define a renormalized coupling and the corresponding RG $\beta$ function,
\begin{equation}
\label{eq:gGF}
g^2_{\GF} (t/a^2;\beta) = \mathcal{N} t^2\langle E \rangle_{t;\beta}\,, \quad\quad \beta(g^2_\text{GF})=- t\frac{d g^2_\text{GF}}{d\,t}\,.
\end{equation}
The expectation value $\langle E \rangle_t$ is calculated on ensembles with fixed bare coupling $\beta=2N_c/g_0^2$ and corresponding lattice spacing $a(\beta)$, so $g^2_{\GF}$  depends both on the lattice flow time $t/a^2$ and $\beta$. Here ${\cal N}$ is a normalization\footnote{If $\mathcal{N} =128\pi^2/(3(N_c^2-1))$, $g^2_{\GF}$ matches the \MSbar{} coupling at tree level \cite{Luscher:2010iy}.} and throughout this work we use a subscript GF to indicate a gradient flow scheme. Only in the $t/a^2 \to \infty$  limit can $g^2_{\GF}$ be interpreted as a renormalized coupling at energy $\mu \sim 1/\sqrt{8t}$. 

Fermionic composite operators evolve according to their scaling dimension under RG. If the RG transformation is linear, as it is the case for the fermion gradient flow \cite{Luscher:2013vga,Luscher:2013cpa}, they also pick up a wave function renormalization factor $Z_\chi$ (corresponding to the $\eta$ exponent in RG language). The vector current has vanishing anomalous dimension, therefore the ratio of any meson operator and the vector current depends only on the anomalous dimension of the meson operator. 

To define our renormalization scheme, we consider a local  flowed operator $\mcO(x;t/a^2)$ at fixed bare gauge coupling $\beta$,
and its correlation function with a probe operator, $\wmcO$, chosen to ensure that the two-point function
\begin{equation}
G_\mcO(x_4;t/a^2,\beta) = a^3\sum_{\bx} \left\langle \mcO(\bx,x_4;t/a^2) \wmcO(0) \right\rangle_{\beta}\,
\label{eq:2p}
\end{equation}
does not vanish. It is natural to chose the unflowed operator   as the probe $\wmcO =\mcO$. 

We define the ratio
\begin{equation}
R_\mcO(x_4;t/a^2,\beta) = 
 \frac{G_\mcO(x_4;t/a^2,\beta)}{G_V(x_4;t/a^2,\beta)}\,,
\label{eq:oR} \end{equation}
with
\begin{equation}
G_V(x_4;t/a^2,\beta) = \frac{1}{3}\sum_{k=1}^3 a^3\sum_{\bx} \left\langle V_k(\bx,x_4;t/a^2) \wV_k(0) \right\rangle_{\beta}\,.
%\label{eq:VV}
\end{equation}
Here $V_k(\bx,x_4;t/a^2)$ is the vector current defined at lattice flow time $t/a^2$ and $\wV$ is an appropriate probe operator. The ratio $R_\mcO(x_4;t/a^2,\beta)$ depends on the anomalous dimension of the operator $\mcO$ and, in the limit $t/a^2 \to \infty$, the logarithmic derivative of $R_\mcO(x_4;t/a^2,\beta)$ predicts the anomalous dimension
\begin{equation}
\gamma_\mcO(t/a^2,\beta) = -2 t 
 \frac{\mathrm{d}\,\text{log}R_\mcO(x_4;t/a^2,\beta)}{\mathrm{d}\,t}\, , \quad\quad  x_4 \gg \sqrt{2Dt} \,.
\label{eq:gammaO_eps}
\end{equation}
The condition $x_4 \gg \sqrt{2Dt}$ ensures that the flowed operator and its probe are well separated, and the correlation function in Eq.~\eqref{eq:2p} decays with an exponent independent of $t/a^2$. In this limit $\gamma_\mcO$ is independent of $x_4$.

The quantity $\gamma_\mcO(t/a^2,\beta) $ could be translated to the energy-dependent running anomalous dimension $\gamma_\mcO(\mu)$, but it is easier and more useful to express $\gamma_\mcO(t/a^2,\beta) $ in terms of the running gauge coupling.  Eq.~\eqref{eq:gGF} gives the running gauge coupling at lattice flow time $t/a^2$ at bare gauge coupling $\beta$, while Eq.~\eqref{eq:gammaO_eps} predicts the anomalous dimension at the same parameters. Combining the two equations therefore predicts $\gamma_\mcO(g^2_{\GF}(t/a^2,\beta))$. The continuum limit corresponds to $t/a^2 \to \infty$ at fixed $g^2_{\GF}$.  That is equivalent to the $\beta \to \infty$ limit, and predicts the running anomalous dimension $\gamma_\mcO(g^2_{\GF})$ in the continuum.
Once the RG anomalous dimensions and $\beta$ function are known, the expression
\begin{equation}
\exp\int_{\bar{g}_\IR}^{\bar{g}_\UV} dg'~\frac{\gamma_\mcO^{\GF}(g')}{\beta^{\GF}(g')}\,
\label{eq:Z_ratio}
\end{equation}
connects the infrared renormalization  factor to its ultraviolet counterpart. 
If we wish to connect with the \MSbar{} scheme, a perturbative calculation has to be carried out.

The equivalence of gradient flow and real-space RG is valid in the infinite volume limit. The definition of the renormalized coupling and RG $\beta$ function in Eq.~\eqref{eq:gGF} and the anomalous dimension in Eq.~\eqref{eq:gammaO_eps} require not only infinite volume but also vanishing fermion mass $am_q=0$. Thus the continuum  renormalization group $\beta$ and $\gamma$ functions are obtained in the $L\to\infty$, $m_q\to 0$, $t/a^2\to \infty$ limit at fixed renormalized coupling $g^2_\text{GF}$.

\section{Preliminary results}
\subsection{Nonperturbative calculation}
\begin{figure}[tb]
\centering
\includegraphics[width=0.48\columnwidth]{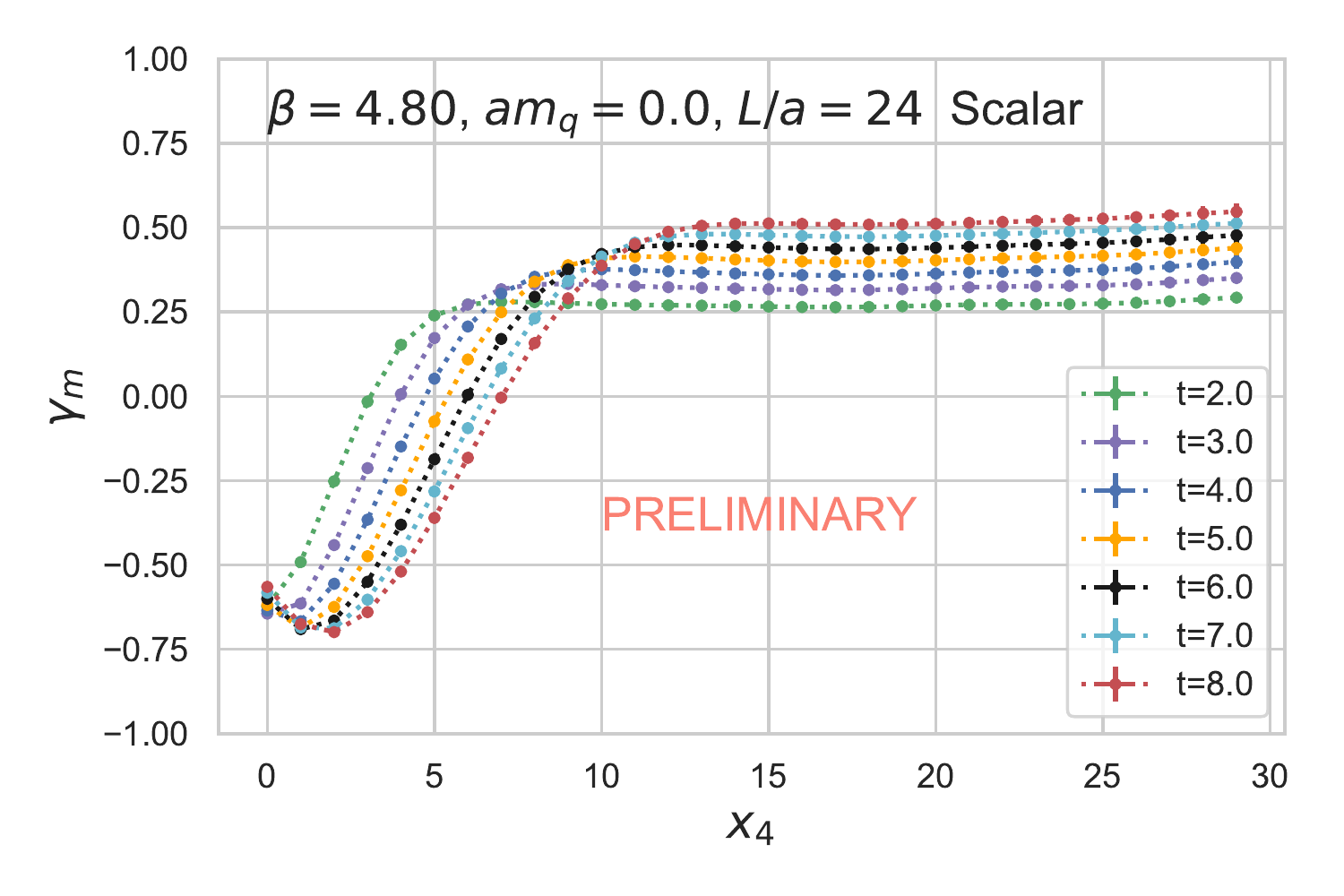} \hfill
\includegraphics[width=0.48\columnwidth]{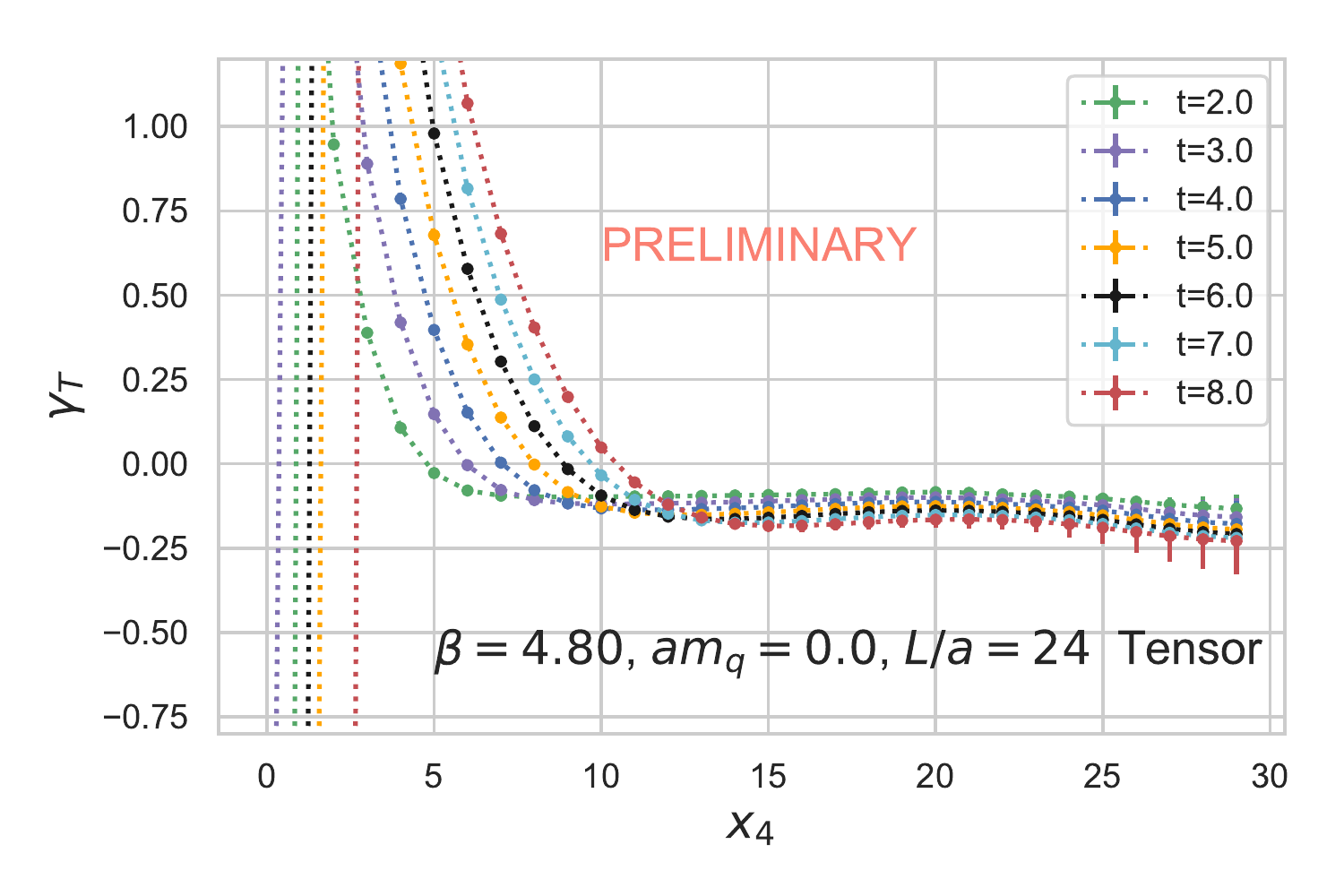}
\caption{The logarithmic derivative of ratio $R_\mcO$, Eqs.~\eqref{eq:oR} and \eqref{eq:gammaO_eps}, as a function of the  separation $x_4$ on the $\beta=4.80$, $am_q=0.0$, $24^3\times 64$ ensemble. The left panel shows the mass (based on the pseudoscalar operator) and the right the tensor anomalous dimensions. In each case seven flow time values  $t/a^2\in(2.0,8.0)$ are shown.}
\label{fig:correlator_ratios}
\end{figure}

\begin{figure}[p] % if not placing both plots on top of each other, change [p] to [tb]
\centering
\includegraphics[width=0.75\columnwidth]{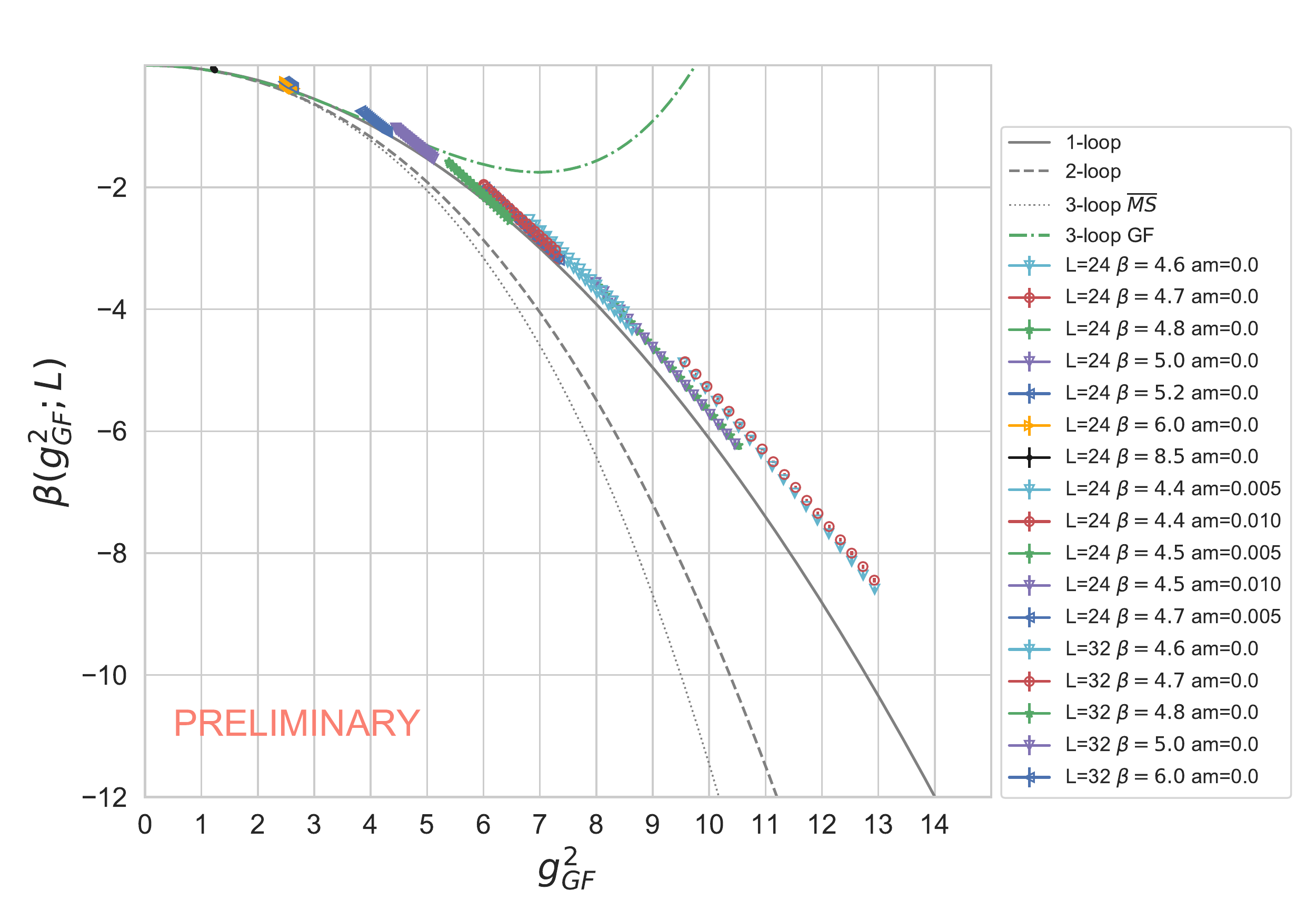} 
\includegraphics[width=0.75\columnwidth]{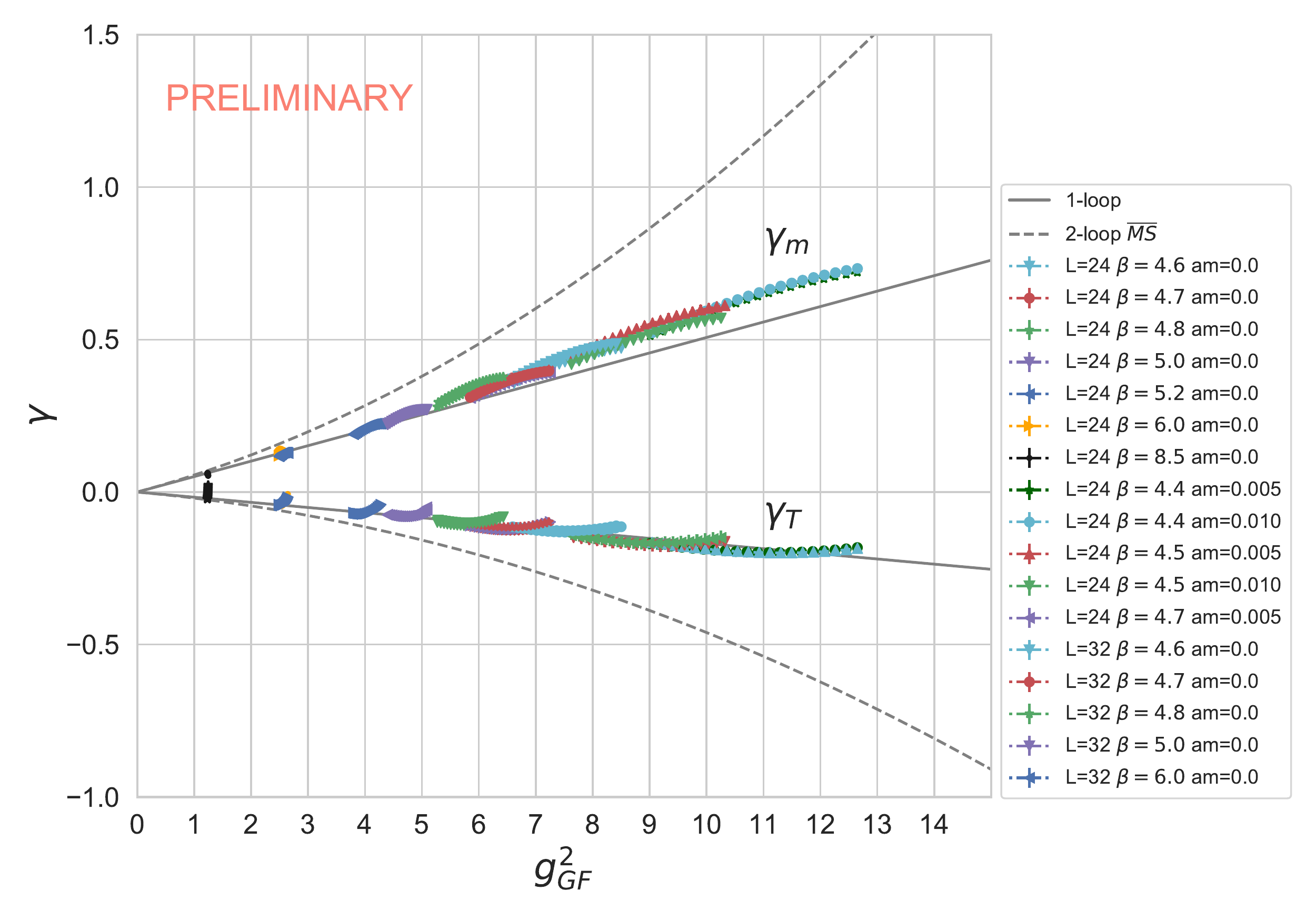} 
\caption{``Raw data'' for the RG $\beta$ function (top panel) and  mass and tensor anomalous dimensions (bottom panel) as a function of $g^2_\text{GF}$ at many different bare coupling $\beta$ values obtained for our $N_f=2$ simulations. On both panels flow times $t /a^2\in (2.0, 4.0)$ are shown on $24^4\times 64$ and $32^3\times 64$ volumes. Simulations in the weak coupling deconfined regime are done in the chiral limit. In the confining regime we use fermion masses $am_q=0.005$ and 0.010. Results in the top panel are based on Wilson flow and Wilson operator. We compare our results to the universal perturbative results at one- and two-loop, as well as three-loop in the \MSbar{} and GF schemes \cite{Harlander:2016vzb}. (For further details see \cite{Hasenfratz:2019hpg}.) In the bottom panel we show the universal one-loop  and the two-loop \MSbar{} predictions. }
\label{fig:raw_vs_g2}
\end{figure}

We demonstrate our new renormalization procedure by performing fermionic gradient flow measurements using a set of gauge field ensembles \cite{Hasenfratz:2019hpg} generated with tree-level improved Symanzik (L\"uscher-Weisz) $SU(3)$ gauge action  \cite{Luscher:1984xn,Luscher:1985zq} and two dynamical flavors of stout-smeared \cite{Morningstar:2003gk} M\"obius domain wall fermions \cite{Brower:2012vk}. In the deconfined region, we perform simulations at zero quark mass and  bare gauge couplings $\beta$ in the range $8.50\ge \beta \ge 4.60$ using volumes  $\left(L/a\right)^3\times T/a = 24^3\times 64$ and $32^3\times 64$. In addition we generated new $24^3\times 64$ ensembles in the confined region with mass $am_q=0.005$ and $0.010$ at $\beta=4.40$ and 4.50, and $am_q=0.005$ at $\beta=4.70$. All configurations are generated using \texttt{GRID}\footnote{\url{https://github.com/paboyle/Grid}} \cite{Boyle:2015tjk} and we implemented our fermionic gradient flow measurements using \texttt{QLUA}\footnote{\url{https://usqcd.lns.mit.edu/w/index.php/QLUA}} \cite{Pochinsky:2008zz}. We choose to use the Wilson kernel with step-size $\epsilon=0.01$ for our gradient flow implementation and apply the Laplace operator to evolve the fermion fields in flow time.

Domain wall fermions preserve chiral symmetry when the fifth dimension $L_s$ is infinite. The explicit symmetry breaking  for finite $L_s$  is characterized by the residual mass $am_\text{res}$. In our simulations we keep $L_s$ large so $am_\text{res} < 10^{-6}$. We find that simulations in the small volume weak coupling regime  preserve chiral symmetry at the level of statistical fluctuations. This ensures, e.g., that the scalar and pseudoscalar anomalous dimensions are identical, $\gamma_\text{S} = \gamma_\text{PS}$.  We have verified this expectation explicitly. 

In Fig.~\ref{fig:correlator_ratios} we illustrate that  $\gamma_\mcO(t/a^2,\beta)$ is independent of $x_4$, the separation between the flowed operator and the probe, if  $x_4 \gg \sqrt{2Dt}$. The left panel  shows the logarithmic derivative defined in Eq.~\eqref{eq:gammaO_eps} for the pseudoscalar operator on the $\beta=4.80$, $a m_q=0.0$, $24^3\times64$ ensemble as the function of $x_4$. The various color symbols correspond to different flow times between $t/a^2=2.0$ and 8.0. While at short distances the flow clearly influences the prediction, all $x_4$ dependence vanishes once $x_4 \gtrsim 2\sqrt{2Dt}$. We use this regime to predict $\gamma_\mcO(t/a^2,\beta)$. 
Figure \ref{fig:correlator_ratios} also illustrates the flow time dependence of $\gamma_\mcO$. For the pseudoscalar operator it shows a steady decrease with $t/a^2$, indicating that the mass anomalous dimension $\gamma_m \equiv \gamma_\text{S}= \gamma_\text{PS}$ increases with decreasing energy. The right panel shows $\gamma_\mcO$ for the tensor operator. Here we observe a negative  anomalous dimension that is  decreasing with increasing $t/a^2$.

Figure \ref{fig:raw_vs_g2} shows our preliminary results for the RG $\beta$ function and for the scalar and tensor anomalous dimensions. Each set of colored symbols correspond to different ensembles and represent the raw data in the flow time range $t/a^2\in(2.0,4.0)$. (Larger flow times correspond to larger $g^2_\text{GF}$ in all cases.) We observe only mild mass dependence. 
The data at $am_q=0.0$, 0.005 and 0.010, when available, are nearly identical. The same is true for the volume dependence. In the investigated flow time range, $L/a=24$ and 32 are barely distinguishable. This implies that the infinite volume and chiral limits will be very similar to the raw data. 
We emphasize that, more importantly, different bare couplings,  covering the range of renormalized couplings $1.0\le g^2_\text{GF} \le 12.0$, map out (approximately) a single curve. Deviations at small $t/a^2$ flow time would imply cut-off effects, whereas deviations at large flow time indicate finite volume effects.  At fixed $g^2_\text{GF}$ values we might have predictions from several different ensembles, i.e.~different $t/a^2$ flow time values. The closeness/overlap of these predictions suggests that the continuum limit  $t/a^2 \to \infty$ extrapolation will also be very similar to the raw data. It is noteworthy that both the scalar and tensor anomalous dimensions lie close to the 1-loop perturbative curve even in the strong coupling range. The $\beta$-function shows similar behavior, pointed out previously in \cite{Hasenfratz:2019puu,Peterson:2021lvb}.

Once the continuum $\beta$ function and anomalous dimensions are obtained, the ratio of the $Z$ factors can be determined according to Eq.~\eqref{eq:Z_ratio}. The final step is the perturbative calculation to connect our gradient flow scheme to the \MSbar{} scheme.

%\begin{figure}[p] % if not placing both plots on top of each other, change [p] to [tb]
%\centering
%\includegraphics[width=0.75\columnwidth]{plots/beta.pdf} 
%\includegraphics[width=0.75\columnwidth]{plots/gamma.pdf} 
%
%\caption{Top: RG $\beta$ function obtained for SU(3) with two fundamental flavors. We compare ``raw data'' based on Wilson flow and Wilson operator obtained from our nonperturbative $N_f=2$ simulations in the deconfined and confined region to universal perturbative results at 1- and 2-loop as well as 3-loop in the \MSbar{} and GF scheme \cite{Harlander:2016vzb}. (For further details see \cite{Hasenfratz:2019hpg}.) Bottom: The mass and tensor anomalous dimensions vs.~$g^2_\text{GF}$ at many different bare coupling $\beta$ values. Flow times $t /a^2\in (2.0, 4.0)$ are shown on $24^4\times 64$ and $32^3\times 64$ volumes. Simulations in the weak coupling deconfined regime are done in the chiral limit. In the confining regime we use fermion masses $am_q=0.005$ and 0.010. }
%\label{fig:gammam_vs_g2}
%\end{figure}

\subsection{Perturbative matching at leading order}
\label{ssec:match}

The matching factor relating our gradient flow scheme and the \MSbar{} scheme can be written in terms of the renormalization group invariant (RGI) operators as
\begin{align}
c^{\ms \leftarrow \GF}(\mu^{\GF}_{\UV},\mu^{\ms}_{\UV}) & = \exp\bigg\{-\int_0^{g_{\GF}^{\UV}} dg'~
\frac{\gamma_\mcO^{\GF}(g')}{\beta^{\GF}(g')} +
\int_0^{g_{\ms}^{UV}} dg'~
\frac{\gamma_\mcO^{\ms}(g')}{\beta^{\ms}(g')}\bigg\}\,.
\label{eq:matching}
\end{align}
Here we denote the renormalized coupling in the $\ms$ scheme by $g_{\ms}$  and the
renormalized coupling in the $\GF$ scheme by $g_{\GF}$.  In general we consider different renormalization scales, 
$\mu^{\GF}_{\UV},\mu^{\ms}_{\UV}$, in the two schemes, both in the perturbative high energy region.

We demonstrate our perturbative matching procedure at leading order in perturbation theory. We determine the next-to-leading-order contributions to the quark bilinear two-point functions, $R_\mcO(x_4;0)$,  by calculating the relevant diagrams at nonzero external momentum transfer and then Fourier transforming the result with respect to the external three momentum. For the leading order anomalous dimension, it is sufficient to determine the flow time dependence of $R_\mcO(x_4;t)$ through a short flow time expansion of renormalized two-point functions \cite{Luscher:2013vga}
\begin{equation}
G_\mcO^R(x_4;\mu^2,t) \sim c_\mcO(\mu^2t)G_\mcO^R(x_4;\mu^2,t=0)  + O(t)\,. 
\end{equation}
This approach has the advantage that the expansion coefficient $c_\mcO(\mu^2t)$ can be extracted from a more straightforward one loop calculation of, e.g., quark bilinears with external quark states.

We need only the divergent contributions to $R_\mcO(x_4;t)$ to extract the leading order anomalous dimension of the quark bilinear $\mcO$. Up to finite terms, the two-point functions are
\begin{equation}
G_\mcO(x_4;t)=G_\mcO^{(0)}(x_4;t)\bigg\{1-\frac{g^2}{12\pi^2}\left[\frac{a_\mcO}{\epsilon} + b_\mcO\log(8\pi\mu^2t) + c_\mcO\log(\pi\mu^2x_4^2)\right] +\mathcal{O}(\epsilon,m,\alpha_S^2)\bigg\},
\end{equation}
where the $a_\mcO$, $b_\mcO$, and $c_\mcO$ depend on the choice of quark bilinear. 

From these results, we evaluate the anomalous dimensions for each quark bilinear through Eq.~\eqref{eq:gammaO_eps}, and find
$\gamma_\mcO = g^2/(12\pi^2) \,  \{6,6,0,0,-2\}+{\cal O}(g^4)$  
for the scalar, pseudoscalar, vector, axialvector, and tensor currents, respectively. These values are in agreement with \cite{Chetyrkin:1997dh,Sint:1998iq} in the cases of the scalar and pseudoscalar currents by comparison with the mass anomalous dimension. A complete next-to-leading order matching calculation will be pursued in future work.

\section{Summary}

We propose a new nonperturbative, gauge-invariant renormalization scheme for local operators in QCD. Our scheme is based on the properties of a real-space renormalization group transformation that we implement through the gradient flow. Here we applied this scheme to local quark bilinear operators, but we expect that the scheme can be implemented for a wide range of local operators. 

We define this gradient flow scheme through a  ratio of two-point functions, which eliminates the wavefunction renormalization of fermion fields at finite gradient flow time.  
The logarithmic derivative of this ratio predicts the  anomalous dimension of the local operator at a given bare coupling $\beta$ and lattice flow time $t/a^2$. When combined with the gradient flow coupling $g^2_\text{GF}$ at the same lattice parameters, we obtain the running anomalous dimension $\gamma_\mcO(g^2_\text{GF})$.  The continuum limit is obtained in the infinite volume chiral limit as $t/a^2 \to \infty$ at fixed $g^2_\text{GF}$. Our numerical data suggest that cutoff effects are small. Curiously, we find that both the RG $\beta$ and $\gamma$ functions are close to their tree-level perturbative values even at strong gauge coupling. In principle, our procedure  maps out the energy dependence of the operator from low to high energies, where the gradient flow scheme can safely be matched to the $\ms$ scheme using perturbation theory. 

The flow time in our scheme introduces a potential ``window problem''.  A clean separation of physics at different length scales requires that the flow time is much larger than the lattice spacing to avoid discretization effects (or, in renormalization group language, to ensure that the flow has reached the vicinity of the renormalized trajectory). At the same time the flow time needs to be much smaller than the volume, to avoid  finite volume contributions. Our preliminary numerical results indicate only mild volume dependence, suggesting that lattice sizes $L/a\ge 24$ are sufficient to carry out the calculation.
%The exact window of validity could be operator dependent and must be studied on a case-by-case basis to optimize the evolution to higher energies through the anomalous dimension. Moreover, f
Future work will be needed to establish the regions of the strong coupling constant for which we can safely perform continuum and infinite volume limits, and to check the stability of the perturbative matching.

\section*{Acknowledgements}

\noindent CJM is supported in part by USDOE grant No.~DE-AC05-06OR23177, under which Jefferson Science Associates, LLC, manages and operates Jefferson Lab. A.H. acknowledges
support by DOE grant DE-SC0010005. MDR and AS acknowledge funding support under the National Science Foundation grant PHY-1913287.

Computations for this work were carried out in part on facilities of the USQCD Collaboration, which are funded by the Office of Science of the U.S.~Department of Energy, the RMACC Summit supercomputer \cite{UCsummit}, which is supported by the National Science Foundation (awards No.~ACI-1532235 and No.~ACI-1532236), the University of Colorado Boulder, and Colorado State University, and the compute cluster \texttt{OMNI} of the University of Siegen. This work used the Extreme Science and Engineering Discovery Environment (XSEDE), which is supported by National Science Foundation grant number ACI-1548562 \cite{xsede} through allocation TG-PHY180005 on the XSEDE resource \texttt{stampede2}.  This research also used resources of the National Energy Research Scientific Computing Center (NERSC), a U.S. Department of Energy Office of Science User Facility operated under Contract  No. DE-AC02-05CH11231.  We thank  Fermilab,  Jefferson Lab, NERSC, the University of Colorado Boulder, the University of Siegen, TACC, the NSF, and the U.S.~DOE for providing the facilities essential for the completion of this work.

\bibliographystyle{JHEP}
\bibliography{proceedings.bib}
\end{document}